\documentclass[journal]{IEEEtran}
\usepackage[dvips]{graphicx}
\usepackage{cite}
\usepackage{paralist}
\usepackage{times}
\usepackage{url}
\usepackage{epsfig}
\usepackage[utf8]{inputenc}
\usepackage{ucs}
\usepackage{pgf}
\usepackage{rotating}
\usepackage{pgf}
\usepackage{subfigure}

\newcommand{\eg}{\textit{e.g.}}
\newcommand{\ie}{\textit{i.e.}}

\begin{document}

\title{On Using P2P Technology for Decentralized Detection of Service Level Agreement Violations}

\author{
\IEEEauthorblockN{J{\'e}ferson C. Nobre, Lisandro Z. Granville} \\
\IEEEauthorblockA{Institute of Informatics, Federal University of Rio Grande do Sul \\
Porto Alegre, Brazil \\
Email: \{jcnobre, granville\}@inf.ufrgs.br} \\
\and
\IEEEauthorblockN{Alberto G. Prieto} \\
\IEEEauthorblockA{Microsoft\\
Milpitas, USA\\
Email: alberto.gonzalez@microsoft.com} \\
\and
\IEEEauthorblockN{Alexander Clemm} \\
\IEEEauthorblockA{Huawei\\
Santa Clara, USA\\
Email: alex@clemm.org}
}

\maketitle

\begin{abstract}

Critical networked services enable significant revenue for network operators and, in turn, are regulated by Service Level Agreements (SLAs). In order to ensure SLAs are being met, service levels need to be monitored. One technique for this involves active measurement mechanisms which employ measurement probes along the network to inject synthetic traffic and compute the network performance. However, these mechanisms are expensive in terms of resources consumption. Thus, these mechanisms usually can cover only a fraction of what could be measured, which can lead to SLA violations being missed. Besides that, the definition of this fraction is a practice done by human administrators, which does not scale well and does not adapt to highly dynamic networking patterns. In this article, we examine the potential benefits of using P2P technology to improve the detection of SLA Violations. We first describe the principles of a P2P-based steering of active measurement mechanisms. These principles are characterized by a high degree of decentralized decision making across a network using a self-organizing overlay. In a second step, we present measurement session activation strategies based on these principles. These strategies do not require human intervention, are adaptive to changes in network conditions, and independent of the underlying active measurement technology.

\end{abstract}


\IEEEpeerreviewmaketitle

\section{Introduction}


Computer network infrastructures have been improving dramatically in terms of capacity and accessibility. Likewise, the communication requirements of distributed services and applications running on top of these infrastructures have become increasingly accurate. Such requirements are usually described in Service Level Agreements (SLAs) established between service providers and customers. To ensure that SLAs are not violated, solutions that allow the service provider to monitor and troubleshoot the underlying communication infrastructure are crucial. 


In SLA monitoring, accuracy and privacy are important aspects. Computer networks are monitored using either active or passive measurement mechanisms, but because active measurement is better than passive in regards to accuracy and privacy, SLA monitoring ends up being strongly based on active mechanisms. In active measurement, probes hosted along the network perform measurement sessions. Synthetic traffic is injected and processed to deliver the network performance in these sessions. Although active measurement is able to detect end-to-end performance problems in a fine-grained way, it is also expensive in terms of the required computing power required to handle activated measurement sessions. As a consequence, active mechanisms consume resources that would be used by primary network functions (\eg, routing and switching). Still, to have a better monitoring coverage, it is necessary to deploy more sessions, which consequently increases resource consumption. On the other hand, observation of just a small subset of all network destinations, to save resources, can lead to an insufficient coverage. Thus, it is crucial to find the sweet spot considering the monitoring needs of each network infrastructure. This tutorial concentrates on active measurements. 


The current practice in activating measurement sessions consists in relying on the network operator's expertise to infer which would be the best network destinations to have sessions activated. Besides of being too labor intensive, this practice is error-prone because even the most experienced operator may identify less than adequate sessions to be activated. In addition, fast changing network conditions also pose challenges, since it is necessary to recompute which sessions should be activated given the new set of critical flows to be observed. Manual, human-centered session activation is certainly not the answer. Management software can be used to help human administrators on the tasks related to the control of active measurement mechanisms. In a first selection, this software can be embedded inside network devices, which is done by some network equipment vendors to avoid the starvation of network devices, \eg, due to configuration errors and lack of experience from junior administrators. However, the manual activation, even aided by current management software, does not enhance the active measurement capabilities in important aspects, such as scalability and efficiency. 


For example, the number of local available measurements results (and, consequently, detected violations) is still bounded by the number of activated measurement sessions. Thus, if the number of SLA violation is greater than the number of available sessions at a given time, only a fraction of the violations will be observed. Also, devices cannot share resources and knowledge about the networking infrastructure in order to take advantage of remote management information (\eg, measurement results). P2P technology can provide the foundations for increasing the intelligence applied in the control of active measurement mechanisms through smart distributed applications. These applications can be deployed in the management software running on the network devices themselves. 


In this article, we examine the potential benefits of using P2P technology to control the deployment of active measurement mechanisms by first reviewing key concepts related to these mechanisms. We then introduce principles to steer autonomically session activation decisions through a self-organizing, embedded P2P measurement overlay which uses the capabilities of network devices themselves. These principle steers a decentralized decision making process: 
\begin{inparaenum}[$i)$]
	\item adaptive to changes in network conditions;
	\item independent of the underlying active measurement technology; and
	\item avoiding human intervention.
\end{inparaenum}
Next, we present strategies to activate active measurement sessions using P2P technology. These strategies highlight the management benefits obtained with our proposed principles, which are summarized, along with concluding remarks and future research opportunities and directions, in the closing section of this article. 

\section{Active Measurement Mechanisms}
\label{active}


Active measurement mechanisms are an important tool to monitor Service Level Objectives (SLO) in network. These mechanisms can be employed in different contexts, such as pre-deployment validation or measurement of in-band live network performance characteristics \cite{IPSLA-Chiba-2013}. Measurements are usually performed by an architecture compromised of measurement probes. These probes inject synthetic traffic into specific network paths to measure the network performance which can lead to either one-way or two-way (\ie, round-trip) measurements. For the sake of simplicity, we will focus on the mechanisms from the Internet Engineering Task Force (IETF) and Cisco Systems.

\begin{figure*}[htb]
  \begin{center}
     \includegraphics[scale=.42]{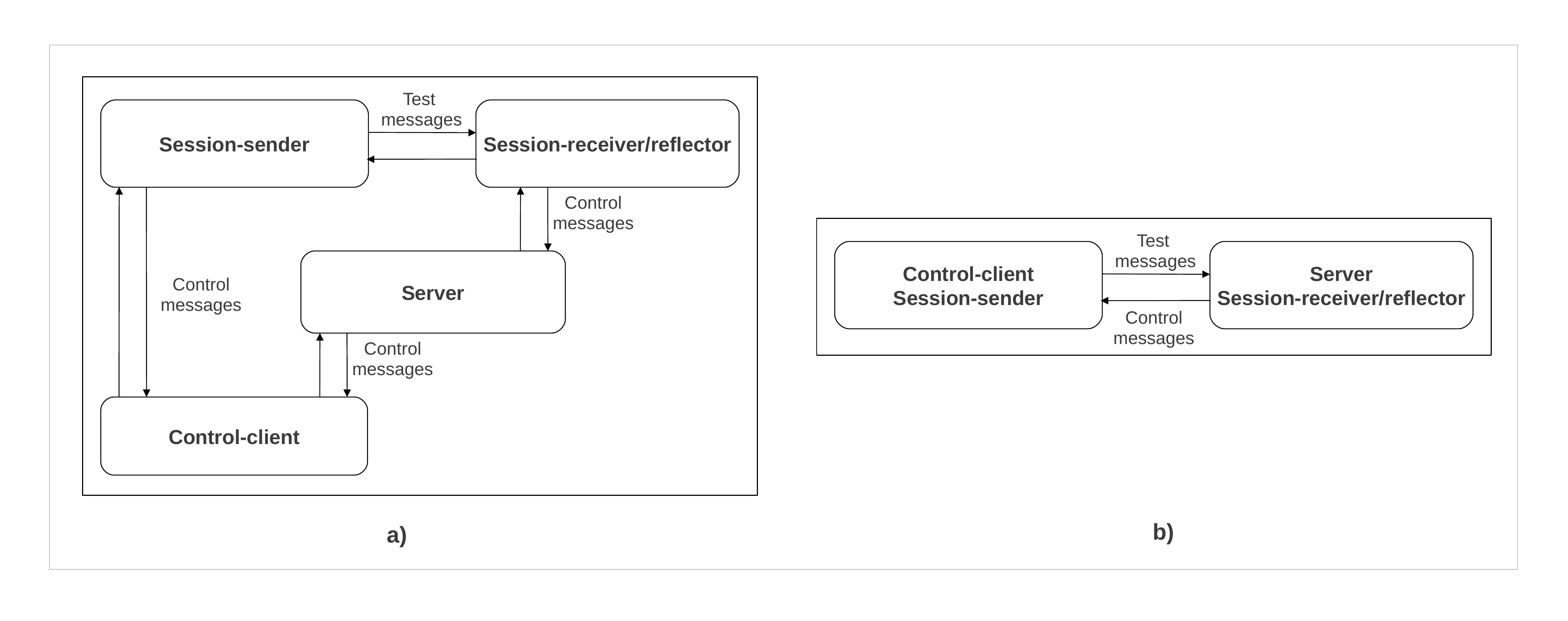}
  \end{center}
  \caption{Active Measurement Logical Models}
  \label{fig:active_model}
	\vspace{-10pt}
\end{figure*}


The IETF IP Performance Metrics (IPPM) Working Group has proposed open active measurement mechanisms that allow the exchange of packets to produce one-way and two-way metrics. These mechanisms are called, respectively, One-way Active Measurement Protocol (OWAMP) \cite{OWAMP-Shalunov-2006} and Two-Way Active Measurement Protocol (TWAMP) \cite{TWAMP-Hedayat-2008}. The O/TWAMP mechanisms consist of 2 inter-related protocols: a control protocol, used to initiate and control measurement sessions and fetch their result, and a test protocol, used to send single measurement packets along the Internet path under test. Control protocol is performed by the control-client (requests, starts, and ends test sessions) and server (manages test sessions); and the test control is the executed by the sender (sending endpoint) and session-receiver/reflector (receiving endpoint). The first part of Figure \ref{fig:active_model} (\textit{a}) shows the logical model used on O/TWAMP considering these roles integrated in 2 end systems. 


Cisco Systems defines the Service Level Assurance (SLA) protocol which is described in an IETF informational RFC \cite{IPSLA-Chiba-2013}. This widely deployed protocol measures service levels related to data link and network layers as well it emulates characteristics of different applications, both considering one-way and two-way metrics. The second part of the Figure \ref{fig:active_model} (\textit{b}) shows the roles described in Cisco SLA protocol, consisting essentially of a sender and a responder. The protocol consists of two distinct phases: the control phase and the measurement phase. The control phase forms the base protocol, which establishes the identity of the sender and provides information for the measurement phase. The measurement phase is comprised of a sequence of measurement-request and measurement-response messages (test messages).


Active measurement mechanisms, such as Cisco SLA protocol and O/TWAMP, have good characteristics for SLA monitoring in terms of accuracy, since they can simulate actual service interactions; and privacy, because they are hard to detect and it is difficult to makes inference about measurements by intermediaries in the middle of the network \cite{OWAMP-Shalunov-2006}. However, these mechanisms are expensive in terms of computational resources consumed on network devices and the network bandwidth required to carry synthetic traffic. The amount of consumed resources is a function of the number of active measurement test sessions, therefore such consumption is directly related to the size and complexity of the network infrastructures. In some settings even dedicated routers (also known as ``shadow routers'') are deployed to handle active measurement mechanisms and save network equipment resources. Thus, an approach to control the deployment of active measurement mechanisms in an autonomic way can be helpful for human administrators.

\section{The Employment of P2P Technology in Network Management}
\label{p2pbnm}


The use of P2P technology in network management, also known as P2P-based Network Management (P2PBNM), extends Distributed Network Management (DNM) by merging characteristics of DNM and P2P overlays \cite{P2PBNM-Granville-2005}. These overlays are known by their scalability, auto-organization, and fault-tolerance. In a P2PBNM system, peers have to perform management tasks and their related provisioning details (\eg, location of peer nodes in the network). From the user perspective, however, the overlay provisioning details must be transparent, requiring no (or little) knowledge about the implementation or architectural organization of nodes in the overlay topology.




The use of local information (data and logic) can improve the execution of management tasks in network infrastructures. Since most of the system state and tasks are directly and dynamically allocated among the peers, management peers do not need to contact centralized management parties. This use also promotes the local autonomy of the peers, incrementing the production of local management decisions. Besides that, P2P management software can be hosted either closer or embedded in network devices, which enables the devices to react faster to changes in management scenarios (\eg, event notification). Moreover, P2P technology can be used to share resources and exchange information among different nodes. 


The federation of management peers can improve their abilities to accomplish tasks. For example, a peer can provide additional resources for other peers when these resources are locally available. In this context, the global ability of the management overlay (\eg, a measurement overlay) can be greater than the sum of the ability of each peer. Sharing resources and exchanging information can also improve the management decisions. Besides local information, management decisions may take into account remote information (\ie, received from other peers) according to the network conditions.

\section{Principles to Steer Measurement Session Activation Decisions using P2P Technology}
\label{principles}

Active measurement mechanisms are an effective technique for monitoring Service Level Objectives (SLOs). In this context, SLA violation detection is based on the idea of identifying deviations from the contracted SLOs. Concerning the utilization of active measurement mechanisms to detect SLA violation, there is an inherent trade-off between attempting to maximize SLA coverage over end-to-end paths and minimizing resource consumption. Intuitively, 2 extreme strategies can be described: maximum coverage, increasing the number of activated measurement sessions without considering resource consumption and possibly leading to network devices starvation; and minimum resource consumption, decreasing network coverage and, probably, missing SLA violations. An effective activation of measurement sessions should balance these strategies. P2P technology can provide the foundations for increasing the intelligence applied in measurement session activation decisions through sophisticated distributed applications.

A pragmatical approach to deploy P2P technology in the control of the activation of active measurement sessions is to define principles to guide this deployment. We devise the utilization of 3 principles:
\begin{inparaenum}[$i)$]
	\item local logic to prioritize destinations using past measurement results;
	\item correlated peers to provision the measurement overlay; and
	\item virtual measurements to optimize resource consumption.
\end{inparaenum}
These principles lead to the introduction of key concepts to support a self-organizing, embedded P2P measurement overlay that uses the capabilities of the network devices to control session activation. In simple terms, our principles try to capture the common sense used by network administrators when using active measurement mechanisms to detect SLA violations. The remaining of the section describes these principles and their implicit concepts.

\subsection{Local Logic for The Destinations Prioritization Using Measurement Results}

The utilization of past service level measurement results is our approach to establish if a destination is likely to disrespect SLOs (\ie, violate the SLA). In order to establish that, we use descriptive statistics metrics to measure the closeness of past service level measurement results regarding the SLO for a given destination. For example, it is possible to use a composition of a measure of the central tendency (\eg, mean) and a measure of spread (\eg, standard deviation) as chosen metrics. If the past measurements results for a given destination are close to a SLO, then the probability of activating a measurement session in this destination should be increased. This is done by local logic, \ie, application that run locally on the network devices.

It is important to assure that each destination is measured frequently, even if its measurement results are not close to the SLOs. In order to induce frequent probing on all destinations, we use the time elapsed from the last measurement for a given destination to increase the probability of this destination to be measured. Clearly, if a destination had not been measured recently, then it should be more likely to be selected in the next measurement decisions. The joint use of the past service level measurement results and the time elapsed from the last measurement for a given destination enables the network devices to determine how to activate sessions in an autonomic manner.

Active measurement mechanisms need to be carefully deployed in order to save computational resources of network devices. Resources utilization on the devices can be managed using constraints. Such constraints are defined according to the maximum number of measurement sessions expected to be deployed in a given time. This number is used as an abstraction for the resources available for active measurement mechanisms. This simplifies the resource management from the device point of view. Since the available resource may vary over time, the constraints should also follow this variation. Since the number of detected SLA violations depends of the active measurement sessions, the rationale for SLA monitoring is to activate sessions while there are available resources. The maximum number of activated measurement sessions can be enforced locally (\ie, in a specific node) and globally (\ie, concerning nodes that exchange management information).

\begin{figure}[tb]
  \begin{center}
     \includegraphics[width=0.45\textwidth]{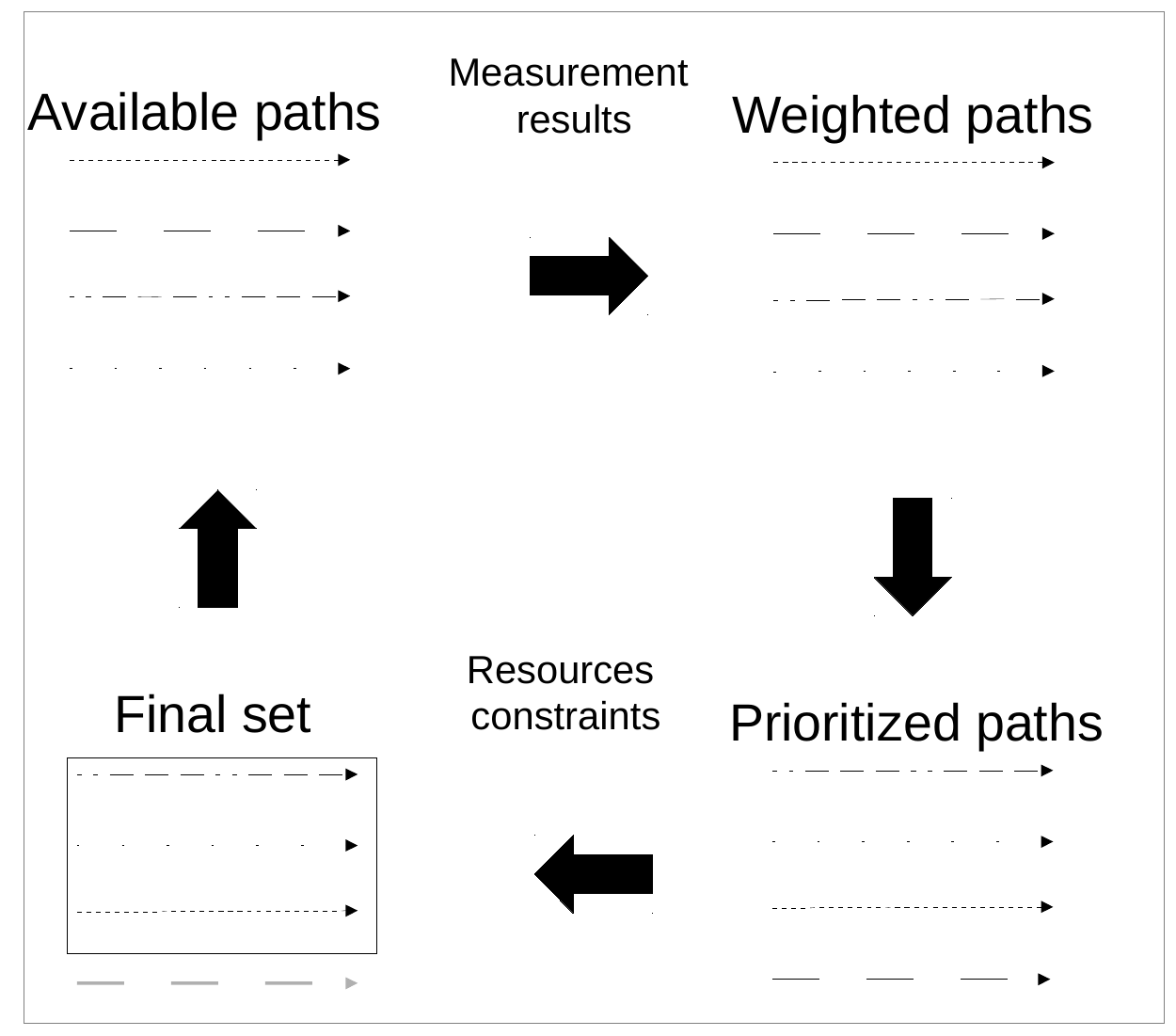}
  \end{center}
  \caption{Phases for the Destination Rank Process.}
  \label{fig:rankloop}
\end{figure}

Destination ranks try to capture the common sense used by network administrators. This common sense is abstracted through destination scores which are used to to steer measurement session activation decisions. The destination rank is composed by the destinations list and their respective scores. The total score of each destination is the sum of its score components. The destination rank process is presented in Figure \ref{fig:rankloop}. The different possible end-to-end destinations for measurement sessions of a specific network device are represented in this figure using arrows. We employ a computationally simple mechanism which is opposite to heavyweight optimization. This mechanism is divided in four phases. First, in the Scores Production phase, destination scores are calculated for the available destinations. Then, these scores are normalized in the Normalization phase. After that, the destinations are prioritized (according to their total score) in the Prioritization phase. Finally, constraints are applied to define the final destination set in Constraints Satisfaction phase. 

%

\subsection{Correlated Peers for Measurement Overlay Provisioning}

Service level measurement results are produced by active measurement mechanisms around the network infrastructure. In this context, human administrators usually can predict if SLA violations are likely to happen in a part of the network infrastructure using information from measurements of other parts of the network. This is possible because administrators can use their experience and knowledge to infer the relation among the links within the network infrastructure. In this context, service level measurements produced by active measurement mechanisms around the network infrastructure could be also shared by the devices to help the local measurement session control. However, it is necessary to assure that the remote results have local relevancy. In order to guarantee that, we use the concept of correlated peers. 2 nodes are considered as correlated peers (correlation is symmetrical) if the results of their measurements for a given destination are correlated. 


For example, if the network paths used by 2 end systems, node \textit{a} and node \textit{b}, to reach a third endpoint, node \textit{c}, are completely disjoint, the contribution due to network transmission for measurement results would be probably different. In this context, the use of measurement results exchanged by the node \textit{a} and node \textit{b} concerning node \textit{c} could lead to undesirable results (\ie, decrease the number of detected SLA violations). In order to determine whether 2 nodes are correlated, we need a substantial amount of past measurements results related to the same destination. Then, it is necessary to compare the local and remote results to verify whether they are in the same vicinity as the local measurements (\eg, low variance). This can be done using correlation functions, such as the Pearson product-moment correlation coefficient and the Spearman rank correlation coefficient.

The use of correlated peers introduces a new topology to deal with: besides network topology (used to find physical neighbors), now there is an overlay topology (the one that that determines which nodes exchange measurement information). To bootstrap such overlay, each network device uses their known endpoints neighbors (presumably from routing or forwarding information bases) as the initial seed to get candidate peers, \ie, peers that may be evaluated for correlation purposes. Then, devices send information about their measurements for their candidate peers. Each device compares this information with their own measurements in order to rank remote devices; top ones are then chosen as correlated peers. Furthermore, peers eventually advertise their correlated peers in order to permit evaluation of ``peers of peers".


\subsection{Virtual Measurements for Resource Consumption Optimization}


Network administrators try to maximize the coverage of a network infrastructure regarding the number of detected SLA violations. However, even considering a na\"{\i}ve attempt of maximum coverage, the number of measurements that a device can perform is still bounded by the available resources (i.e., the number of measurement sessions which a device can actually deploy considering their resource consumption). Sharing active measurement results among devices can improve SLA violation detection regarding the resource consumption and monitoring coverage. This sharing can be done by devices agreeing to exchange measurement results and contracting measurement session activation. P2P technology can be used to enable resources sharing and information exchanging among a measurement overlay. 


Our proposed solution to increase the number of SLA violation tries to capture one of the behaviors commonly employed by network administrators, the sharing of measurement results. Sometimes a single device cannot achieve the desired measurement coverage in isolation due to its own capabilities. Besides that, the administrator can choose not to achieve a defined coverage considering the device in isolation. This is usually done to save resources for main network functions, such as switching and routing. However, it is important to define which network devices are prone to share measurement results, considering their own capabilities, the quality constraints, and the available resources. We define a virtual measurement session as the use of results from remote measurement sessions by a peer as their own.

\begin{figure}
 \begin{center}
\includegraphics[width=0.45\textwidth]{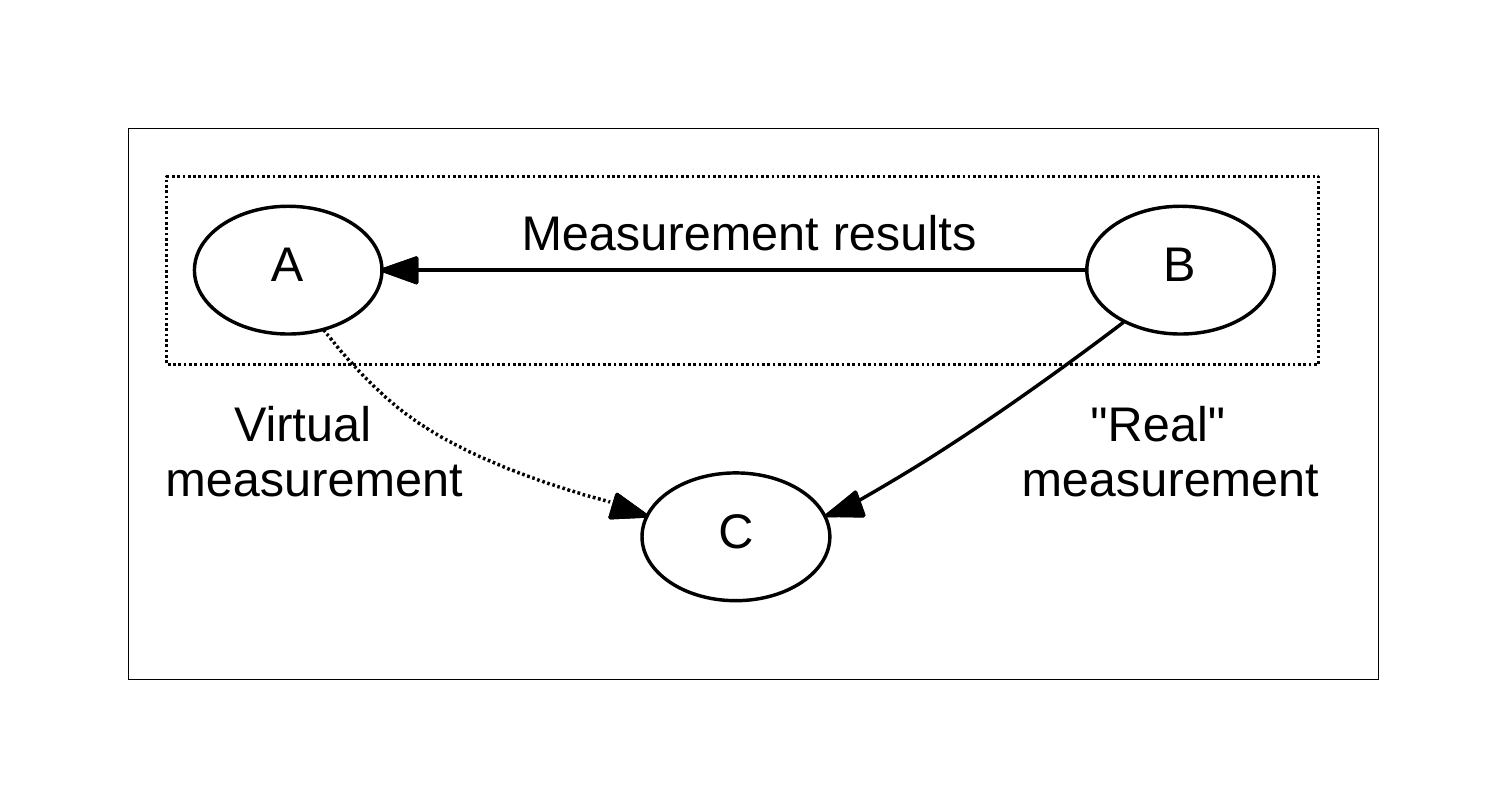}
 \end{center}

 \caption{Virtual Measurement Session.}
  \label{fig:virtual}
\end{figure}

The Figure \ref{fig:virtual} shows a virtual measurement session. In this figure, device \textit{b} performs a measurement session using device \textit{c} as destination. Then, device \textit{b} sends measurement results produce in this measurement session to device \textit{a} which, in its turn, uses these results as its own results, i.e., a virtual measurement session. The resource consumption due to the traffic injection and packet handling concerning this example is carried by device \textit{b} since it is the one actually performing the measurement session. Device \textit{a} enjoy the use of (virtual) measurement results to device \textit{c}, avoiding the measurement session overhead.

 
The concept of correlated peers can be used to assist the definition of virtual measurement sessions. The devices decide which peers are the best choices to share measurement sessions using their own list of correlated peers. Since the measurement overlay is built using information from past service level measurement results, the choice of virtual sessions is indirectly related with the same results. Furthermore, the P2P measurement overlay is responsible to forward the results of remote measurement sessions.

\section{Strategies to Activate Measurement Sessions}

The principles presented in Section \ref{principles} are defined to steer autonomically measurement session activation decisions. In order to accomplish these principles we define measurement session activation strategies, which are the expected behavior from the network devices concerning the control of active measurement mechanisms. We assume these mechanisms must be controlled (as described in the strategies) without internal modification, \ie, the strategies should be able to handle current versions of active measurement mechanisms. Thus, the strategies aim at increasing the efficiency of the detection of SLA violations solely through efficient measurement session activation decisions. This assumption increases the applicability of the present work.

Initially, we define 3 strategies to choose which destinations will be probed: measurement session activation based solely on local information, measurement session activation based on both local and remote information, and the use of virtual measurement sessions using the measurement overlay. Each strategy builds up on the previous one, increasing the used information. A priori, as more information from the network is used, the measurement session activation decisions capture better the service level violations.

\begin{table*}[!htbp]
\centering
\caption{Summary of the Proposed Measurement Session Activation Strategies.}
\label{table:strategies}
\begin{tabular}{ |l|l|l|l|l|l|l| }
  \hline
  Strategy &  \multicolumn{3}{|c|}{Principles} & \multicolumn{3}{|c|}{Criteria} \\
  \hline
{}   & Local Logic   & Correlated Peers & Virtual Measurement & Adaptivity & Scalability & Accuracy\\
  \hline
Local   &  Yes & No   & No  & Low & Low & High\\
Local and remote &  Yes & No & No  & High & Low & High\\
Virtual  &  Yes  &  Yes   & Yes & High & High & Low\\
  \hline
\end{tabular}
\end{table*}

In Table \ref{table:strategies}, we provide a summary of measurement session activation strategies according to their adherence to the principles presented in Section \ref{principles} and a qualitative evaluation on some features (Adaptivity, Scalability, and Accuracy). For an experimental evaluations such strategies, see \cite{P2PBNM-Nobre-2012,P2PBNM-Nobre-2013,P2PBNM-Nobre-2014}. The remaining of the section presents the description of each one of the measurement session activation strategies.

\subsection{Local Strategy}


The local strategy is performed using only information locally available on a node. This information comes from past service level measurement results. This information is used to compute the scores for each destination. As the local strategy is the simplest one, only 2 scores are used. The first one tries to capture that destinations which are closer to violate the SLA should have a higher probability of being probed in the following iterations. In more detail, the score uses the average of past measurement results (discounted over time) considering a sliding window. The second score aims at maintaining frequent measurements on destinations. Thus if a destination had not been measured recently, then it should be more likely to be selected in the next iterations. As the measurement results are updated dynamically, the strategy can adapt to changes in network conditions using the destination rank.

Despite the use of local logic and data, the local strategy uses only a few desirable features of P2P technology. Each device has its own destination rank which leads to a distributed destination rank among the network infrastructure. Such distribution promotes the local autonomy of the devices, incrementing the production of local management decisions. However, since the local strategy uses only information locally available, just the local upper bound for the number of activated measurement session must be observed, thus there is not a significant improvement in the SLA detection scalability. In any case, even the solely use of locally collected measurement results and the observation of the local upper bound lead to better results than the random activation in terms of the number of SLA violation detections and the adaptivity to changes in network conditions \cite{P2PBNM-Nobre-2012}.

\subsection{Local and Remote Strategy}


The local and remote strategy is performed using information available from the local network device and received from other devices. Thus, now the destination rank contains scores related to remote information. Therefore, the main difference between this strategy and the local one is the source of measurement results. Now, in each iteration, there are 2 distinguished phases: peer topology phase and measurement session activation. In the first phase, we use the concept of correlated peers (as described in Section \ref{principles}). After that, the measurement session activation takes place using the locally collected information and also measurement results received from correlated peers.


The local and remote strategy uses a measurement overlay built using correlated peers. In order to define its correlated peers, a device needs past measurement results from other devices. Each round is preceded by peer selection, which determine the set of candidate peers to share measurement results. After the first interaction, current peers can be used for re-evaluation. After determine actual correlated peers from candidate peers, the device starts to send its past measurement results for the correlated peers. As a note, peer topology updates and probing may run on separate schedules; they do not have to adhere to the same time intervals.


Some constraints are necessary to enable an efficient overlay operation for the network-wide control of active measurement mechanisms. The use of these constraints is twofold: they define which is the minimum measurement correlation to set devices as correlated peers and how many peers a device can have. The minimum correlation score required for a correlated peer sets a lower bound constraint. This is necessary to assure local relevance of remote results. On the other hand, the maximum number of correlated peers of a peer in a given time is controlled by a upper bound constraint. Since resources are needed for peer maintenance, the upper bound restricts the resource consumption.

The measurement session activation phase takes place in a similar fashion that in local strategy. However, the destination score also takes into account the past measurement results from correlated peers. This can improve the adaptivity of the measurement session activation decision, since SLA violations detected in different parts of the network can steer the activation of local sessions. The local and remote strategy does not use virtual measurement sessions, thus the detection accuracy is analogous to the one found on the local strategy.

\subsection{Virtual Strategy}

The virtual strategy enhances the concept of correlated peers in order to choose which peers are interesting to share measurement sessions with, \ie,  to have virtual measurement sessions with. The bulk of the computation, bandwidth, and storage needed to operate the virtual measurement session is contributed by the correlated peer. The virtual strategy builds up on the local and remote strategy, thus the peer topology and peer probing phases are also performed (as explained before). In this context, the adaptivity of the virtual strategy and the local and remote strategy are similar. Peers partition the SLA monitoring tasks (\ie, available destinations) for the set of destinations they are interested in and have no measurements. We consider a scenario of multiple devices which observe multiple events (active measurement sessions) and those devices need to perform measurement session decisions in a dynamic network. Hence some mechanism is needed to dynamically adapt the contract of virtual measurement sessions to network conditions. 


Peers employ a computationally simple algorithm to contract measurements from other devices through message exchange in a P2P measurement overlay. In order to accomplish this contract, we have developed a simple protocol \cite{P2PBNM-Nobre-2013}. In order to start the virtual measurement session, a coordination request is sent by the device to the chosen correlated peer. The peers that receive the coordination request can return the request with either a positive coordination response or a negative coordination response. If the devices receives a negative coordination response or the device is allowed to have more virtual measurement sessions, the strategy will continuously try to find other correlated peers to deploy such sessions. Correlated peers can request and inform the stop of measurement tasks regarding different destinations.


The virtual strategy can decrease the local resources necessary to monitor destinations, thus improving the scalability of the SLA monitoring process. On the one hand, the use of virtual measurement sessions increases the possible number of detected SLA violations which can be higher than the number of ``real" activated measurement sessions. On the other hand, results from virtual sessions probably are not as accurate as if they had been produced locally. Thus, virtual measurement sessions can result in false positives and negatives regarding the detection of SLA violations. In order to improve virtual session accuracy, only top correlated peers can be used, \ie, peers with high correlation. We assume that higher correlation between peers are an indicative of trustiness in the virtual measurement sessions. This can enhance the confidence among peers, which is important to ensure that the virtual strategy leads to desirable results. Finally, devices can also control the resource consumption due to virtual measurement sessions using resources constraints.

\section{Final Remarks and Future Directions}

Critical networked services established between service provider and customer are expected to operate respecting Service Level Agreements (SLAs). An interesting possibility to monitor these SLAs is using active measurement mechanisms. However, these mechanisms are expensive in terms of resources consumption and also increase the network load because of the injected traffic. In addition, if the number of SLA violation in a given time is higher than the number of available measurement measurement sessions (common place in large and complex network infrastructures), certainly some SLA violations will be missed.  The current best practice, the observation of just a subset of network destinations driven by human administrators' expertise, is error prone, does not scale well, and is ineffective on dynamic network conditions. This can lead to SLA violations being missed, which invariably affects the performance of several applications. In this article, we advocated the use of P2P technology to increase the potential number of detected SLA violations.  


We proposed 3 principles to steer management decisions regarding active measurement mechanisms: the use of local logic to prioritize destinations using past measurement results, correlated peers to provision the measurement overlay, and virtual measurements to optimize resource consumption. These principles are accomplished through strategies to activate measurement sessions. These strategies are based on the utilization of P2P-Based Network Management (P2PBNM) features (\textit{e.g.}, high degree of decentralization) to enable the sharing of measurements among network devices and the exchange of information to improve management decisions.


Our work is intended to be an initial step towards P2P-based control of active management mechanisms. We also intend to investigate how coordination features can enable composite measurement tasks, \ie,  measurements can be break down into sub-tasks, and, thus lead to the cooperation of peers among the P2P measurement overlay to further increase the number of detected SLA violations. Besides that, refinements in the definition of correlated peers can be included to allow a more selective peering. For example, throttling of overlay traffic can be introduced for ``popular" peers. Furthermore, there will likely be other interesting research opportunities in the context of present work. For example, information about correlated peers can have other uses, \eg, allow inferences about the underlying (physical) topology. 


\bibliographystyle{IEEEtran}
\bibliography{jcnobre-SLAcontrol-tutorial}

\end{document}